\documentstyle[preprint,aps,psfig]{revtex}
\begin{document}
\draft
\preprint{HUPD-9821}
\title{Pattern of Chiral Symmetry Restoration at Finite Temperature\\
       in A Supersymmetric Composite Model}
\author{J. Hashida, T. Muta and K. Ohkura}
\address{Department of Physics\\
         Hiroshima University\\
         Higashi-Hiroshima, Hiroshima 739-8526}
\date{\today}
\maketitle
\begin{abstract}

The structure of chiral symmetry restorations at finite temperature is
thoroughly investigated in the supersymmetric Nambu-Jona-Lasinio model
with a soft supersymmetry breaking term. It is found that the broken
chiral symmetry at vanishing temperature is restored at sufficiently
high temperature in two patterns, i. e., the first order and second order
phase transition depending on the choice of the coupling constant $G$
and the supersymmetry breaking parameter $\Delta$. The critical curves
expressing the phase boundaries in the $G-\Delta$ plane are completely
determined and the dynamically generated fermion mass is calculated
as a function of temperature.

\end{abstract}


\narrowtext

The supersymmetry is an essential ingredient of unified
field theories for elementary particles so that it is worth
trying to study a supersymmetric version of composite
Higgs models such as Technicolor model. We shall consider
a supersymmetric composite model in the present communication
with an interest in the early universe and take into account
finite temperature effects.
As a prototype of the composite model we pick up the
Nambu-Jona-Lasinio (NJL) model which is known to be useful
to investigate the mechanism of the 
dynamical symmetry breaking \cite{Nam}.
The supersymmetric version of the NJL model is easily constructed.
Unfortunately, however, in the supersymmetric NJL model
the chiral symmetry is strongly protected 
to keep the boson-fermion supersymmetry (SUSY) and hence
the dynamical chiral symmetry breaking does not take place
\cite{Lov}.
If a soft SUSY breaking term is added to 
the SUSY NJL Lagrangian, the dynamical breakdown of the 
chiral symmetry is brought about for sufficiently large 
SUSY breaking parameter $\Delta$ \cite{Ell}.
The reason for this is simple: The large $\Delta$ implies the large effective 
mass of the scalar components of the superfields so that quantum effects due
to the scalar components get suppressed compared with that of the spinor
components. Thus the model becomes closer to the original NJL model which
allows the dynamical fermion mass generation.
Stating the same substance in a different way
we realize that the boson by acquiring its mass term forces supersymmetry to 
make balance so that the fermion mass is generated dynamically.

The SUSY NJL model with soft SUSY breaking is useful to study the mechanism of
the dynamical chiral symmetry breaking within the framework of supersymmetry.
If we take the model seriously as a prototype of the unified field
theory, it is natural to extend the argument to take into account circumstances
of the finite temperature and spacetime-curvature as in the early universe.
It should, however, be noted that SUSY is not a good symmetry for finite
temperature \cite{Oji}. This fact may discourage studies of SUSY models
at finite temperature. In our argument we consider the supersymmetry of the 
model in the strict sense only at vanishing temperature and focus 
our attention on the dynamical breaking of the chiral symmetry in the same
model at finite temperature.
The breaking of SUSY at finite temperature is kinematical since it is caused
essencially by the difference of the statistical distribution functions for 
bosons and fermions.

     A natural SUSY extension of the NJL model is characterized by 
the following Lagrangian \cite{Lov},

\begin{eqnarray}
{\cal L}_{SNJL}=\int d^4\theta && 
    [\bar{\Phi}_{+}\Phi_{+}+\bar{\Phi}_{-}\Phi_{-}
    +\frac{G}{N}\ \Phi_{+}\Phi_{-}\bar{\Phi}_{+}\bar{\Phi}_{-}
       \nonumber \\ 
    &&  -\Delta^2
    \theta^2 \bar{\theta}^2(\bar{\Phi}_{+}\Phi_{+}+\bar{\Phi}_{-}\Phi_{-})],
    \label{Lsnjl}
\end{eqnarray}
where $d^4 \theta=d^2\theta d^2\bar{\theta}$. The chiral superfield 
${\Phi}_{\pm}$
is represented by component scalar ($\phi_{\pm}$) and
spinor ($\psi_{\pm}$) fields and auxiliary field $F_{\pm}$ so that
$\Phi_{\pm}(y)=\phi_{\pm}(y)+\sqrt{2}\theta\psi_{\pm}(y)+F_{\pm}(y)\theta^2$ 
with $y=x+i\theta\sigma^\mu\bar{\theta}$. The chiral superfields
$\Phi_{+}$ and $\Phi_{-}$ belong to the multiplet {$N$} and {$\bar{N}$} in
$SU(N)$ respectively. Note that a soft SUSY breaking term is introduced in
Eq. (\ref{Lsnjl}) following Buchmueller and Ellwanger \cite{Ell}.

In the following arguments it is convenient to introduce auxiliary chiral
superfields $H$ and $S$ and use a Lagrangian equivalent to 
Eq. (\ref{Lsnjl})

\begin{eqnarray}
{\cal L}
&=&\int d^4\theta [(\bar{\Phi}_{+}\Phi_{+}+\bar{\Phi}_{-}\Phi_{-})
  (1-\Delta^2\theta^2 \bar{\theta}^2)+\frac{N}{G}\bar{H}H]
\nonumber \\
&+&[\int d^2\theta\ (\frac{N}{G}HS-S\Phi_{+}\Phi_{-})+h.c.].
\end{eqnarray}
Written in terms of component fields the Lagrangian takes the following
form where we have kept only relevant terms for calculating our effective 
potential in the leading order of the $1/N$ expansion:
\begin{eqnarray}
{\cal L}&=&-\bar{\phi}_{+}(\Box +|\phi_S|^2+\Delta^2)\phi_{+}
    -\bar{\phi}_{-}(\Box +|\phi_S|^2+\Delta^2)\phi_{-} \nonumber\\
&&  +\bar{\Psi}(i\partial_\mu\gamma^\mu-\sigma+i\gamma_5\pi)\Psi
-\frac{N}{G}|\phi_S|^2.\label{Laux}
\end{eqnarray}
The field $\phi_S$ is related to the ordinary scalar ($\sigma$) 
and pseudoscalar
($\pi$) field such that $\phi_S=\sigma + i \pi$, 
and $\Psi=(\psi_+,\bar{\psi}_-)^T$.

     We would like to calculate the effective potential for Lagrangian
(\ref{Laux}) in the leading order of the $1/N$ expansion at finite 
temperature.
We adopt the Matsubara formalism \cite{Mat} to introduce temperature effects
in our calculation of the effective potential.
We simply replace the integral in energy variable by the summation where we
apply the periodic boundary condition for boson fields and the anti-periodic
boundary condition for fermion fields respectively.
The calculation of the effective potential is essentially the same as
the one described in our previous paper \cite{Ina}.
The summation can be easily performed and after some algebra the following 
result is obtained,
\begin{eqnarray}
V(|\phi_S|^2)=V_0(|\phi_S|^2)+V_\beta(|\phi_S|^2),\label{V}
\end{eqnarray}
where $V_0(|\phi_S|^2)$ is the effective potential for vanishing temperature,

\begin{eqnarray}
V_0(|\phi_S|^2)&=&\frac{|\phi_S|^2}{G}+2\int\frac{d^{D-1}k}{(2\pi)^{D-1}}
    (\ \sqrt{k^2+|\phi_S|^2+\Delta^2}\nonumber\\&&-\sqrt{k^2+|\phi_S|^2}
    -\sqrt{k^2+\Delta^2}+\sqrt{k^2}\ ),\label{V0}
\end{eqnarray}
which agrees with the one obtained by performing the $k_0$ integration after
Wick rotation in the effective potential given at vanishing temperature, and
$V_\beta(|\phi_S|^2)$ is the temperature dependent part of the effective 
potential
that is given by

\begin{eqnarray}
V_\beta(|\phi_S|^2)&=&-\frac{4}{\beta}\int\frac{d^{D-1}k}{(2\pi)^{D-1}}
  \ln(\frac{1+e^{-\beta\sqrt{k^2+|\phi_S|^2}}}{1+e^{-\beta\sqrt{k^2}}}
\nonumber\\&&
  \hspace{1cm}\times\frac{1-e^{-\beta\sqrt{k^2+\Delta^2}}}
  {1-e^{-\beta\sqrt{k^2+|\phi_S|^2+\Delta^2}}}),\label{Vbeta}
\end{eqnarray}
where $D$ is the space-time dimension (Although $D=4$, we keep $D$ to be 
arbitrary
in Eqs. (\ref{V0}) and (\ref{Vbeta}) for later convenience),
$\beta$ is the inverse of the temperature $T$ in unit of
$k_B=1$ with $k_B$ the Boltzmann constant.
Note that the effective potential (\ref{V}) is normalized so that $V(0)=0$.

     It is easy to see that the effective potential (\ref{V}) reduces
to $V_0(|\phi_S|^2)$ in the zero temperature limit $\beta\to \infty $. 
If we let
$\Delta\to 0$ in the zero temperature effective potential (\ref{V0}), 
we are left
only with the tree term as is expected by the boson-fermion cancellation in
the SUSY model.
In the SUSY limit $\Delta\to 0$ the finite temperature effective
potential (\ref{Vbeta}) still retains the term of quantum corrections which 
is
composed of the bosonic and fermionic statistical distribution functions.
This result reflects the well-known fact that supersymmetry is not a good
symmetry at finite temperature \cite{Oji}.
At sufficiently low temperature
both the Fermi-Dirac and Bose-Einstein distribution
function $1/(\exp {\beta\sqrt{k^2+|\phi_S|^2}\pm 1})$ are well approximated
by the Maxwell-Boltzmann distribution function 
$\exp (-\beta\sqrt{k^2+|\phi_S|^2})$
and hence the quantum correction term in Eq. (\ref{V}) disappears.
Thus supersymmetry is a good symmetry not only at zero temperature 
but also in
the low temperature region.

     Although the phase structure of our model at finite temperature
is studied by direct numerical estimates of the effective potential 
(\ref{V}), it is more convenient to deal with the gap equation
$V'(|\phi_S|^2)=0$ where the prime denotes the differenciation with respect to
$|\phi_S|^2$ : $V'(|\phi_S|^2)=\partial V/\partial |\phi_S|^2$,

\begin{eqnarray}
\frac{1}{G}&-&\int\frac{d^{D-1}k}{(2\pi)^{D-1}}
(\frac{1}{\sqrt{k^2+|\phi_S|^2}}-\frac{1}{\sqrt{k^2+|\phi_S|^2+\Delta^2}})
\nonumber\\
&&      +2\int\frac{d^{D-1}k}{(2\pi)^{D-1}}[\ \frac{1}{\sqrt{k^2+|\phi_S|^2}}
\frac{1}{e^{\beta\sqrt{k^2+|\phi_S|^2}}+1}\nonumber\\
&&      -\frac{1}{\sqrt{k^2+|\phi_S|^2+\Delta^2}}
        \frac{1}{e^{\beta\sqrt{k^2+|\phi_S|^2+\Delta^2}}-1}\ ]=0.\label{Gap}
\end{eqnarray}

     In order to see the situation as clearly as possible 
we start with the discussion in the case $D=3$
where the gap equation (\ref{Gap}) after performing the integration
is much simpler
than the one for $D=4$ and there is no divergence in the integration in
Eq. (\ref{Gap}). The gap equation for $D=3$ is obtained 
in an analytical form by performing the integration in Eq. (\ref{Gap}):

\begin{eqnarray}
\frac{2\pi}{G}-\sqrt{|\phi_S|^2+\Delta^2}+\sqrt{|\phi_S|^2}+\frac{2}{\beta} 
\ln {\frac{1+e^{-\beta\sqrt{|\phi_S|^2}}}
{1-e^{-\beta\sqrt{|\phi_S|^2+\Delta^2}}}}=0.
\label{Gap3d}
\end{eqnarray}
By numerical inspections we find that Eq. (\ref{Gap3d}) allows at most two
nontrivial solutions for $|\phi_S|^2$ depending on the choice 
of parameters $G$, 
$\Delta$ and $\beta$. Thus the effective potential (\ref{V}) 
with $D=3$ may have
zero, one and two extremum (extrema) respectively. Hence, when the temperature
$1/\beta$ is increased, we will have possibilities of experiencing no phase
transition, the 2nd order phase transition and the 1st order phase transition.

     The effective potential $V(|\phi_S|^2)$ for $D=3$ is calculated
analytically such that
\begin{eqnarray}
V(|\phi_S|^2)&=&\frac{|\phi_S|^2}{G}+\frac{1}{3\pi}[|\phi_S|^3+\Delta^3-
(|\phi_S|^2+\Delta^2)^{3/2}] \nonumber \\&&\hspace{-0.5cm}
+V_{\beta}(|\phi_S|^2)_{D=3},\label{V3d}
\end{eqnarray}
where we have suppressed writing explicitly the temperature dependent term
(the last term on the right hand side of Eq. (\ref{V3d})) which is
rather lengthy and can be expressed by using the polylogarithms.

At vanishing temperature the chiral symmetry in the model is broken \cite{Ell}
if sufficiently large is the parameter $\Delta$ appearing in the soft SUSY
breaking term. In fact by observing the behavior of the gap equation
$V_0'(|\phi_S|^2)=0$ we can confirm the statement.
In three dimensions for vanishing temperature $\beta\to \infty$ the last term
on the left hand side of Eq. (\ref{Gap3d}) disappears and we see that only 
one
nontrivial solution for $|\phi_S|^2$ exists when $2\pi/G < \Delta$.
The curve on the $G-\Delta$ plane given by the equation
\begin{eqnarray}
G\Delta/(2\pi)=1 \label{Crit3d0}
\end{eqnarray}
divides the whole $G-\Delta$ plane into two regions which represent the unbroken
and broken chiral symmetry respectively.

If the parameters $G$ and $\Delta$ are kept in the unbroken region, we remain
in the unbroken region when temperature is increased. If the parameters
$G$ and $\Delta$ are in the broken region, we experience the chiral symmetry
restoration as temperature gets high enough. This situation is easily
confirmed by observing the gap equation (\ref{Gap3d}) and the effective 
potential
(\ref{V3d}). In the present situation there are two types of the chiral 
symmetry
restoration : the first-order and second-order phase transition. These two cases
can be distinguished from each other by the conditions $V'(0)=0, V''(0)<0$
and $V'(0)=0, V''(0)>0$ respectively. Hence the condition that $|\phi_S|^2=0$
be a point of inflection for the effective potential, $V'(0)=V''(0)=0$, divides
these two cases and gives a curve on the $G-\Delta$ plane.
The curve is the critical curve on the $G-\Delta$ plane which divides the region 
of
the broken phase into the regions of the first and second order phase transition.
Written in parameters $G$ and $\Delta$ the condition $V'(0)=V''(0)=0$ reads
\begin{eqnarray}
&&\frac{2\pi}{G\Delta}=1-\frac{2}{\beta\Delta}
\ln{\frac{2}{1-e^{-\beta\Delta}}}, \label{Crit3d1a}\\
&&\beta\Delta/2=\coth (\beta\Delta/2) \label{Crit3d1b}.
\end{eqnarray}
Here $\beta\Delta/2$ is determined by Eq. (\ref{Crit3d1b})
and is equal to 1.19968.
The critical curve distinguishing the second order phase transition
from the first order one is given by the equation obtained
by substituting the above value for $\beta\Delta/2$
in Eq. (\ref{Crit3d1a}), i.e.,
\begin{eqnarray}
G\Delta/(2\pi)=2.91631. \label{Crit3d1}
\end{eqnarray}
The whole $G-\Delta$ plane is divided by two hyperbolas (\ref{Crit3d0})
and (\ref{Crit3d1}) into three regions which correspond to the
unbroken, broken (2nd order) and broken (1st order) region respectively as
shown in Fig. 1.
Note here that as can be seen in Eq. (\ref{V3d}) the effective potential for
$D=3$ reduces to that of the ordinary NJL model by letting $\Delta\to \infty $.
The physical reason for this result is simple: In the limit of large $\Delta$
the scalar components of the superfield carry a large
effective mass and decouple from the spinor components leaving only the
contribution of the NJL fermions. It should, however, be remarked
that the gap equation (\ref{Gap3d}) diverges as $\Delta\to \infty$. This 
result
corresponds exactly to the fact that the gap equation for the ordinary
NJL model diverges linearly for $D=3$. In fact it is given by
\begin{eqnarray}
\frac{2\pi}{G}=\Lambda-\sqrt{|\phi_S|^2}-
\frac{2}{\beta}\ln(1+e^{-\beta\sqrt{|\phi_S|^2}})
\end{eqnarray}
and corresponds to Eq. (\ref{Gap3d}) with large $\Delta$ replaced by $\
Lambda$.
In the ordinary NJL model for three dimensions the bare coupling constant $G$ is
to be kept small for large $\Lambda$. This fact precisely reflects the property
that $G$ gets smaller as $\Delta$ becomes larger along the critical curve.

     We then go over to the case of the four dimensional space-time.
For $D=4$ the situation is almost the same as in $D=3$. The main difference
comes from the presence of the divergence for $D=4$ in the integration of
Eqs.(\ref{V}) and (\ref{Gap}). We regularize the divergence by
introducing the three dimensional cut-off $\Lambda$ in the momentum integration.
The effective potential for $D=4$ at vanishing temperature $V_0(|\phi_S|^2)$ is
calculated analytically while the temperature dependent part of the effective
potential $V_\beta(|\phi_S|^2)$ is estimated numerically. In these calculations
all the parameters in the model are scaled by the cut-off $\Lambda$.
By direct numerical estimates we find that the effective potential (\ref{V})
with $D=4$ represents the chiral symmetry restoration as temperature gets high
enough. In Fig. 2 (a) and (b) we see that the chiral symmetry restoration is of
the 2nd as well as 1st order phase transition depending on the choice of the
parameters $G$ and $\Delta$ (In our previous paper the existance of the 2nd 
order
phase transition was overlooked \cite{Has}).

The gap equation $\partial V/\partial |\phi_S|^2=0$ reads
\begin{eqnarray}
\frac{1}{2\alpha}&-&\sqrt{1+x}+\sqrt{1+x+\delta}
    -(x+\delta)\ln\frac{1+\sqrt{1+x+\delta}}{\sqrt{x+\delta}}\nonumber\\
    &&+x \ln\frac{1+\sqrt{1+x}}{\sqrt{x}}
    +\frac{4\pi^2}{\Lambda^2}\frac{\partial V_\beta(|\phi_S|^2)_{D=4}}
    {\partial |\phi_S|^2}=0,\label{Gap4d}
\end{eqnarray}
where $\alpha$, $\delta$ and $x$ are defined by $\alpha=G\Lambda^2/(8\pi^2),
\delta=\Delta^2/\Lambda^2, x=|\phi_S|^2/\Lambda^2$.
For $\beta\to\infty$ the last term on the left hand side in Eq. (\ref{Gap4d})
drops out leaving only temperature independent terms and we have the gap 
equation
at zero temperature. It is not difficult to show that Eq. (\ref{Gap4d}) 
without
the temperature dependent term allows a nontrivial solution for
$x$ when the parameters $\alpha$ and $\delta$ satisfy the condition,

\begin{eqnarray}
\frac{1}{2\alpha} < 1-\sqrt{1+\delta}+\delta\ln
    \frac{1+\sqrt{1+\delta}}{\sqrt{\delta}}.\label{Crit4d0}
\end{eqnarray}
Thus the dynamical fermion mass is generated and the chiral symmetry is broken
dynamically at vanishing temperature if the above inequality is satisfied.
The boundary of the region given by
Eq. (\ref{Crit4d0}) in the $\alpha-\delta$ plane is the critical 
curve dividing the symmetric and broken phase as shown in Fig. 3.

     Just in parallel with the case of $D=3$ we find three possibilities in 
which the gap equation (\ref{Gap4d}) allows no solution, one 
nontrivial solution and two nontrivial solutions respectively
depending on the choice of parameters $G$, $\Delta$ and $\beta$.
It is easily found by observing the behaviors of the gap equation and
the effective potential that there are two types of phase transitions,
the first order and second order, as temperature increases.
As in the case of $D=3$ the critical curve distinguishiding
the region of the 1st order phase transition from the 
region of the 2nd order phase transition is given by solving
simultaneous equations for $G$ and $\Delta$ corresponding to
the following equations,
\begin{eqnarray}
 V'(0)=V''(0)=0. \label{Crit4d1}
\end{eqnarray}
The critical curve derived from Eq. (\ref{Crit4d1}) is shown by
a dotted line in Fig. 3.

The full and dotted line divide the whole $G-\Delta$ region into three regions
of the unbroken chiral symmetry, the broken chiral symmetry with the 2nd order
phase transition and the broken chiral symmetry with the 1st order phase
transition respectively.
The dynamical fermion mass generated by the above chiral symmetry breaking
is obtained by numerically solving the gap equation (\ref{Gap4d})
as is shown in Fig. 4.

One of the direct physical consequences of our approach may be found in
the field of electroweak baryogenesis in early Universe \cite{Cli}.
In this connection it is interesting to see whether composite Higgs masses
can be taken large enough to account for the present experimental
observation and to construct realistic composite Higgs models
for describing the baryogenesis \cite{Boc}.
As a physicsl application of our approach we may consider the
supersymmetric $E_6$ GUT model \cite{Kug} in which the symmetry breaking 
patterns can be precisely investigated by using our results.
It is also interesting to note whether our results presented in the paper is
subject to any change if we introduce the contribution of
order $1/N$ or higher.
Such calculation has been done in Ref. \cite{Esp} in the case of the
nonsupersymmetric NJL model in three dimensions and an extension of
their result to
the case of the SUSY NJL model should be straightforward.
The investigation in this direction will be left for future works.

After the completion of our work we became aware of a report discussing the 
similar analysis as in the present paper \cite{Lal}.  The conclusion of
Ref. \cite{Lal}, however, is quite different from ours.

The authors would like to thank K. Kikkawa and C. S. Lim for useful
comments on supersymmetry at finite temperature and T. Inagaki,
S. Mukaigawa and M. Tanabashi for enlightening discussions.

\begin{figure}[htbp]
  \begin{center}
    \centerline{\psfig{figure=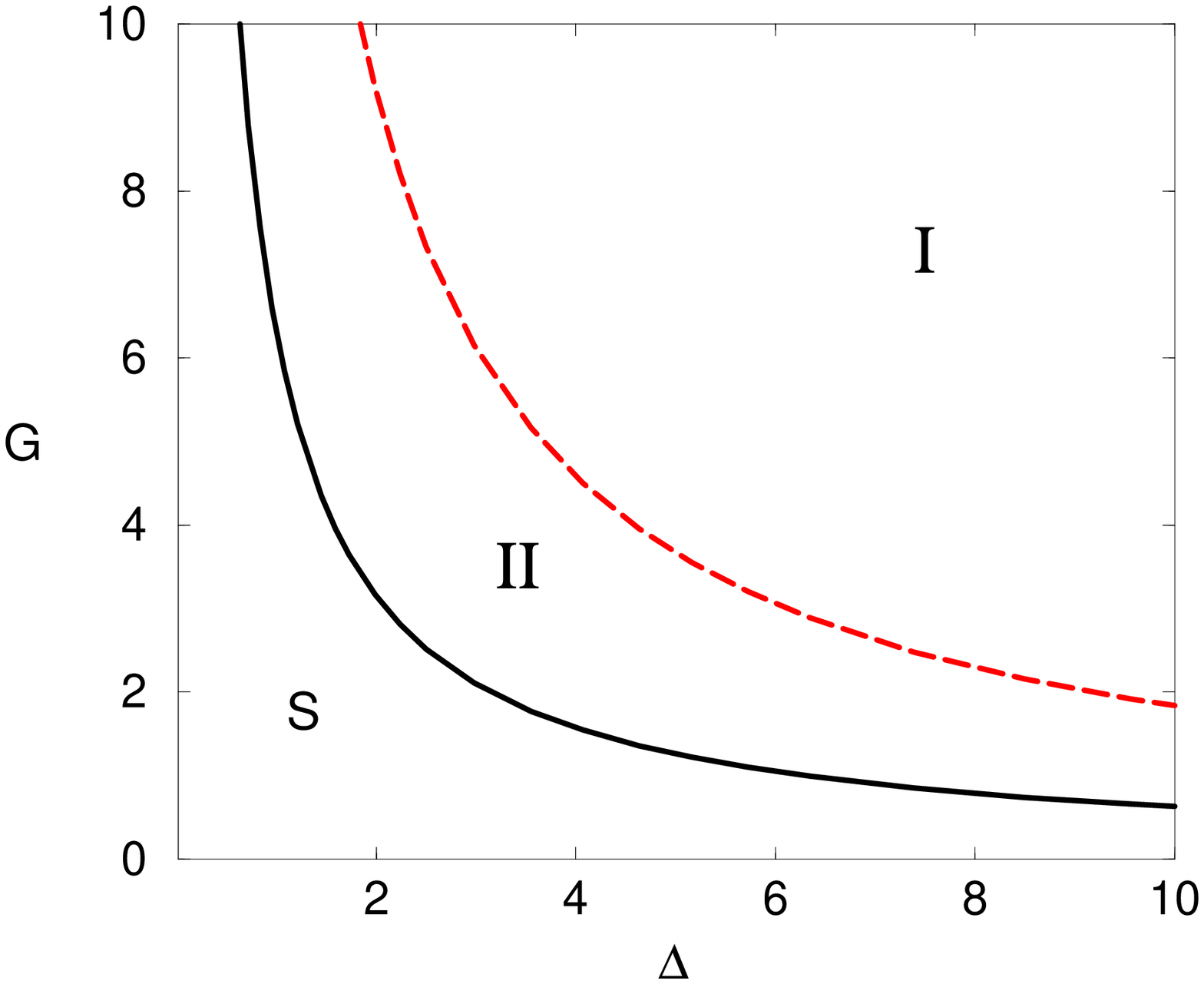,width=7cm}}
        \caption{Critical curves on $G-\Delta$ plane in $D=3$: 'S' represents
            the region of the parameters $G$ and $\Delta$ in which the chiral
            symmetry is not broken at any temperature, 'II' represents
            the region of the parameters in which the broken
            chiral symmetry at vanishing temperature is restored at high
            temperature as the second order phase transition and
            'I' represents the region of the parameters
            in which the broken chiral symmetry at vanishing temperature is
            restored at high temperature as the first order phase transition.}
    \label{potential1}
  \end{center}
\end{figure}

\begin{figure}[htbp]
  \begin{center}
    \centerline{\psfig{figure=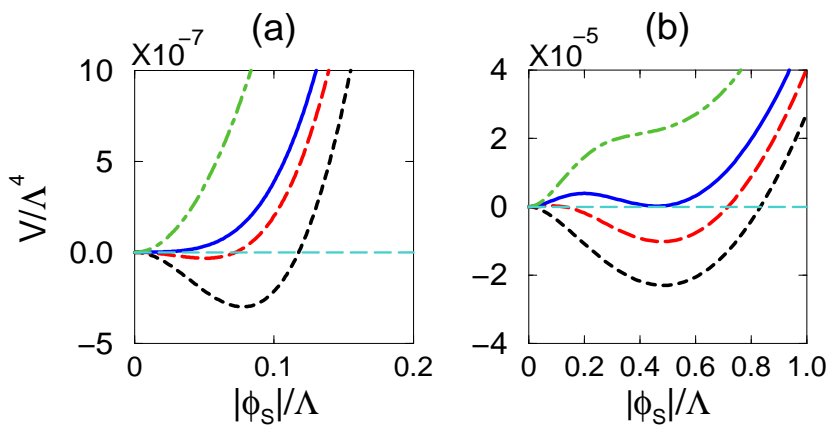,width=12cm}}
          \caption{The $\beta$ dependence of the effective 
          potential (a) for $\alpha=25$
          and $\delta=0.1$ and (b) for $\alpha=90$ and $\delta=0.1$.}
    \label{potential1}
  \end{center}
\end{figure}

\begin{figure}[htbp]
  \begin{center}
    \centerline{\psfig{figure=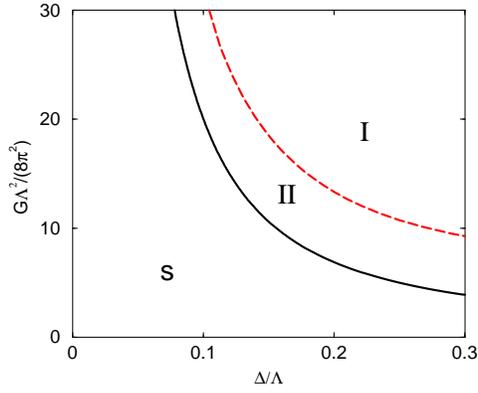,width=7cm}}
    \caption{Critical curves on $G-\Delta$ plane in $D=4$: The explanation of
             symbols 'S', 'II' and 'I' is the same as the one in Fig. 1.}
    \label{potential2}
  \end{center}
\end{figure}

\begin{figure}[htbp]
  \begin{center}
    \centerline{\psfig{figure=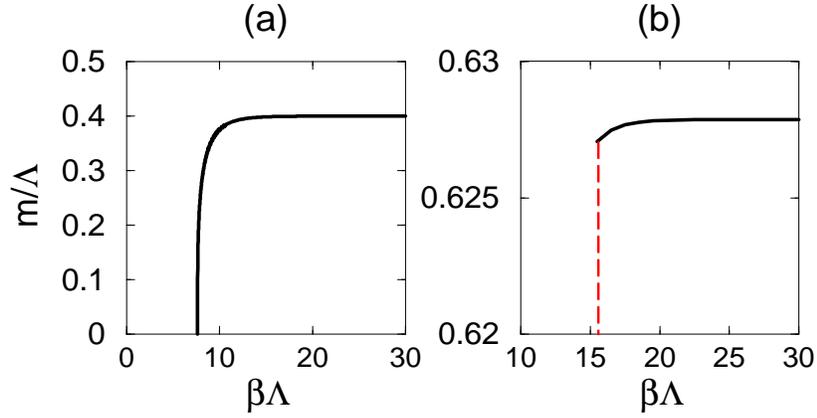,width=12cm}}
    \caption{The dynamical fermion mass as a function of $\beta$ (a) for
             $\alpha=25$ and $\delta=0.1$ and (b) for $\alpha=90$ and
             $\delta=0.1$.}
    \label{potential3}
  \end{center}
\end{figure}

\end{document}